\documentclass[epj]{svjour}
\usepackage{graphics}
\begin{document}
\title{Use of the chemical potential for a limited number of fermions with a
degenerate groundstate}
\titlerunning{N-Fermions and the chemical potential}
\author{L.F. Lemmens\inst{1} \and F. Brosens\inst{2} \and J.T. Devreese\inst{2}
}                     
%
%
\institute{Departement Natuurkunde, Universiteit Antwerpen (RUCA),\\
Groenenborgerlaan 171, B-2020 Antwerpen, Belgium. \and Departement Natuurkunde, Universiteit Antwerpen (UIA),\\
Universiteitsplein 1, B-2610 Antwerpen, Belgium,}
\date{Received: date / Revised version: date}
%
\abstract{
For fermions with degenerate single-particle energy levels, the usual
relation between the total number of particles and the chemical potential $%
\mu $ is only satisfied for a specific number of particles, i.e. those
leading to closed shells. The treatment of an arbitrary number of fermions
requires a modification of the chemical potential, similar to the one
proposed by Landsberg for Bose-condensed systems. We study the implications
of the required modification for fermions in a  potential, by
calculating the ground state energy, the free energy, the density, the
partition function and the dynamic two-point correlation function. It turns
out that the modified relation between the fugacity and the number of
particles leads to the correct ground state energy and density. But for
other quantities like the entropy and the two-point correlation functions,
an additional correction is required and derived. These calculations
indicate that many-body perturbation theories based on $H-\mu N$ with
Lagrange multiplier $\mu $, are not applicable in unmodified form for a
fixed number of fermions at low temperature.
} 
\maketitle
\section{Introduction}
\label{intro}
When studying many-body effects in systems with Hamiltonian $H$ for a finite
number  $N$ of particles, it is often convenient to introduce the chemical
potential $\mu $ as a Lagrange multiplier associated with $N$  \cite%
{AGD.AP63,FW.MGH71,NO.PB85}, and to investigate properties related to $H-\mu
N$. E.g., the Gibbs free energy $G$ is obtained from $e^{-\beta G}=%
\mathop{\rm Tr}%
e^{-\beta \left( H-\mu N\right) },$ where the trace is taken over all the
degrees of freedom, {\em including} the number of particles. After the
calculations, the chemical potential is then determined from the constraint
that the number of particles under consideration be $N$, e.g., to obtain the
free energy $F$ from $e^{-\beta F}=%
\mathop{\rm Tr}%
e^{-\beta H},$ where the latter trace is taken over the degrees of freedom
of the $N$ particles, with $N$ fixed. This procedure can provide a very
efficient way to deal with the finite particle-number, but it has to be
handled with care  \cite{SW.PRA98}, because in unmodified form it is only
valid for closed shell systems.

We show that a modified expression for the fugacity $u$ of the form $%
u=\lambda e^{\mu \beta }$, as introduced by Landsberg  \cite{L:PR54} in the
case of Bose condensed systems, instead of the usual expression $e^{\mu
\beta }$, provides a relation between the total number of particles and the
fugacity for any number of particles, also if the ground state is
degenerate. It leads to the correct ground state energy and density. But
even the use of the modified constraint on the number of particles does not
provide correct results for other quantities. E.g. the calculations of the
entropy and of the two-point correlation functions require specific and
well-defined corrections, to be derived below.\label{0}

Our derivation is based on the $N$-particle density matrix, properly
anti-symmetrized for fermions. With the aid of this quantity we can express
the partition function, the density and the density-density correlation
function in terms of single-particle density matrices. The method allows for
the analytical calculation of the partition function and the correlation
functions in the low temperature limit. We show explicitly how the partition
function and the density can be calculated in order to elucidate the same
method used in the more involved calculation of the density-density
correlation function. In essence, the method proceeds as follows. The
anti-symmetrization induces a summation over all possible cycles of the
permutation group. In order to avoid the difficult constraint on the cycle
lengths, arising from the fixed number of particles, a generating function
is constructed by forming a power series in $u$ with the elements $A_{N}$ of
the sequence $\left\{ A_{N}|N=1,\ldots ,\infty \right\} $ as coefficients,
where $A_{N}$ is the result of the cyclic summation for $N$ particles. This
method is standard in quantum statistical mechanics  \cite%
{K.NH74,F.B72,F.PR53,M.PTP51,PO.PR56}, and very useful to construct path
integrals for identical particles  \cite%
{LVV.PRA93,C.RMP95,BDL.PRE97a,BDL.PRE98a,VVIG.PRE99}. 
Replacing $u$ by $\frac{1}{z}$ it is indicated as the $Z$-transform: 
$$ \widetilde{A}(z)=\sum_{n=0}^{\infty}A_{n}\frac{1}{z^{n}}$$

Once the generating functions are known we have to expand them in a power
series in $u,$ and extract the coefficient of the $N^{th}$ power of $u$ to
obtain the expression corresponding with a fermion gas of $N$ particles  \cite%
{EG.PRA86}. Using this technique the partition function, the one-point and
two-point correlation function can be obtained directly i.e. without the use
of any quantity related to the chemical potential.

It should be noted that we could easily have advertised this investigation
under the common denominator of ``non-equivalence between ensembles''  \cite%
{ZUK.PRp77,FHW.JSP70,BDL.SSC96,GH.PRL97,NBGIR.PRL97,GR.PRL97,IGNWR.PRL99,HK.AnP99,R.PRC95,P.PRL00,MS:PRL00}%
. But we refrain from doing so for the following reason. Starting from a
stationary Hamiltonian with a countable spectrum, one can construct a
density of states for a given total energy and a given number of particles.
The partition function is the Laplace transform of this density of states
with respect to the energy, and the grand canonical partition function is
the generating function of this partition function with respect to number of
particles. In this respect the three functions only describe the {\em same}
information in different parameter spaces $$\left\{ E,N\right\}
\underbrace{\Longleftrightarrow}_{\rm Laplace-transform} \left\{ \beta ,N\right\}
\underbrace{\Longleftrightarrow}_{\rm Z-transform} \left\{ \beta ,u\right\}. $$ Only if
the transforms are {\em approximately} evaluated, the warranty that they
lead to the same results expires  \cite{LBDSSC99}. This approach is based on
the fundamental transforms between $\Omega $, $Z$ and $\Xi ,$ and it
indicates that the calculation of properties of a system with a finite
number of particles using the chemical potential -- even with the
modification -- is an approximation as soon as the levels of the
Fermi-system are degenerate. 

In the next section we introduce the basic methods, and illustrate them by
analyzing the partition function. In the subsequent sections, the same
methodology is used to study the one-point and the two-point correlation
functions. Finally, we will discuss our results and their implications for
the $N$-body problem.
\section{The partition function of an open shell fermion gas}

In this section we will elucidate the mathematical methods used throughout
the paper, by analyzing the partition function of an open shell fermion gas
as an example. It is well known that the Fermi-Dirac distribution can be
derived from the probabilistic problem of distributing $N$ fermions on a
given set of states, assuming that a state can contain at most one fermion.
The resulting distribution is easily derived using generating functions  \cite%
{W.PB70}. Assume that the energy of a state is given by $\epsilon _{\nu }$.
The energy $E$ of the ground state (i.e., the total energy of all occupied
states at zero temperature) can then be calculated. In the case of a
one-dimensional (1D) oscillator with frequency $\omega $ the energy spectrum
is given by $\epsilon _{\nu }=\hbar \omega \left( \nu +\frac{1}{2}\right) ,$
with degeneracy $g_{\nu }=1.$ Placing $N$ fermions in the system leads to
the ground state energy $E=\hbar \omega \sum_{\nu =0}^{N-1}\left( \nu +\frac{%
1}{2}\right) .$ The occupancy density $p_{\nu }$ of a state expressed in
terms of its energy becomes $p_{\nu }=\frac{1}{N}\Theta \left( \epsilon
_{\nu }<\epsilon _{N}\right) ,$ where $\epsilon _{N}$ is the lowest energy
of the unoccupied states. The generalized Heaviside function used here is
defined as $$\Theta \left( x\right) =\left\{ 
\begin{array}{cc}
1 & \rm{if }\quad x\quad \rm{ is \, true} \\ 
0 & \rm{otherwise}.%
\end{array}%
\right. $$ However, the non-degenerate case is the exception rather than the
rule. We might also consider an illustration of the case of open shells.
E.g., for a two-dimensional (2D) oscillator, the number of states with the
same energy $\epsilon _{\nu }$ is characterized by the degeneracy $g_{\nu
}=\nu +1,$ whereas for the 3D oscillator $g_{\nu }=\left( \nu +1\right)
\left( \nu +2\right) /2.$ The set of states with the same energy $\epsilon
_{\nu }$ is called a level or a shell, labeled by the (nonnegative integer)
index $\nu $.

If $L$ denotes the index of the lowest not completely occupied shell (i.e.,
either empty or partially filled), then the number $N_{L}$ of fermions
accommodated in completely filled shells is $N_{L}=\sum_{\nu =0}^{L-1}g_{\nu
}.$ If not all shells are completely filled, there are $N-N_{L}$ fermions
left over to occupy the shell with index $L$. The ground state energy is
therefore 
\begin{equation}
E_{0}(N)=\sum_{\nu =0}^{L-1}g_{\nu }\epsilon _{\nu }+\left( N-N_{L}\right)
\epsilon _{L}.
\end{equation}%
The energy difference at low temperature between a gas containing $N$
particles and one containing $(N-1)$ particles is given by: 
\begin{equation}
E_{0}(N)-E_{0}(N-1)=\epsilon _{L},
\end{equation}%
provided that the gas with $N$ particles is an open shell gas. In textbooks
this difference is sometimes used as the definition of the chemical
potential. In the present paper, we follow the convention that $\mu $ is the
parameter that is used to satisfy the constraint arising of the finite
number of particles as discussed in the introduction.

\subsection{From density of states to generating function and vice versa}

The local density of states matrix $\Omega \left( E,{\bf r}^{\prime },{\bf r}%
\right) $ for a particle in a one-body potential can be defined as 
\begin{equation}
\Omega \left( E,{\bf r}^{\prime },{\bf r}\right) =\sum_{\nu =0}^{\infty
}\delta \left( E-\epsilon _{\nu }\right) \sum_{k=1}^{g_{\nu }}\varphi _{\nu
,k}\left( {\bf r}^{\prime }\right) \varphi _{\nu ,k}^{\ast }\left( {\bf r}%
\right) ,  \label{eq:Omegamatrix}
\end{equation}%
where $\varphi _{\nu ,k}\left( {\bf r}\right) $ are ortho-normal wave
functions with corresponding eigenvalues $\epsilon _{\nu }.$ The index $\nu $
accounts for the energy levels and the index $k$ runs from $1$ to $g_{\nu },$
where $g_{\nu }$ is the degeneracy of the energy level $\epsilon _{\nu }.$
The diagonal part of this matrix is given by 
\begin{equation}
\Omega \left( E,{\bf r}\right) =\sum_{\nu =0}^{\infty }\delta \left(
E-\epsilon _{\nu }\right) \sum_{k=1}^{g_{\nu }}\left| \varphi _{\nu
,k}\left( {\bf r}\right) \right| ^{2},  \label{eq:Omegadiag}
\end{equation}%
and the density of states $\Omega \left( E\right) $ becomes 
\begin{equation}
\Omega \left( E\right) \equiv \int \Omega \left( E,{\bf r}\right) d{\bf r}%
=\sum_{\nu =0}^{\infty }g_{\nu }\delta \left( E-\epsilon _{\nu }\right) .
\end{equation}

The single-particle propagator  in its spectral representation is usually written as 
\begin{equation}
K_{1}\left( {\bf r}^{\prime },\beta |{\bf r},0\right) =\sum_{\nu =0}^{\infty
}e^{-\beta \epsilon _{\nu }}\sum_{k=1}^{g_{\nu }}\varphi _{\nu ,k}\left( 
{\bf r}^{\prime }\right) \varphi _{\nu ,k}^{\ast }\left( {\bf r}\right) ,
\end{equation}%
and it is clearly the Laplace transform of $\Omega \left( E,{\bf r}^{\prime
},{\bf r}\right) $%
\begin{equation}
K_{1}\left( {\bf r}^{\prime },\beta |{\bf r},0\right) =\int \Omega \left( E,%
{\bf r}^{\prime },{\bf r}\right) e^{-\beta E}dE,
\end{equation}%
with $\beta $ adjoined to $E,$ provided the Laplace transform exists, i.e.,
that the energy levels $\epsilon _{\nu }$ are positive. This means that the
origin of the energy spectrum is chosen such that it does not exceed the
lowest eigen-energy $\epsilon _{0}$. The relation between the single-particle
partition function $Z_{1}\left( \beta \right) $ and the single-particle
propagator 
\[
Z_{1}\left( \beta \right) =\int \int K_{1}\left( {\bf r}^{\prime },\beta |%
{\bf r},0\right) \delta \left( {\bf r}^{\prime }-{\bf r}\right) d{\bf r}d%
{\bf r}^{\prime }, 
\]%
implies that the single-particle partition function can also be written as a
Laplace transform of the density of states: 
\begin{equation}
Z_{1}\left( \beta \right) =\sum_{\nu =0}^{\infty }g_{\nu }e^{-\beta \epsilon
_{\nu }}=\int \Omega \left( E\right) e^{-\beta E}dE,  \label{eq:Z1spectral}
\end{equation}%
with the inverse relation given by 
\begin{equation}
\Omega \left( E\right) =\frac{1}{2\pi i}\oint_{c}Z_{1}\left( s\right)
e^{sE}ds,  \label{eq:invOMZ}
\end{equation}%
where $c$ is a Bromwich contour.

The propagator ${\cal K}_{D}$ of $N$ distinguishable particles with position
vectors ${\bf r}_{j}$ $\left( j=1,\ldots ,N\right) $, all in the same
potential, is the product of their single-particle propagators 
\begin{equation}
{\cal K}_{D}\left( \bar{r},\beta ;\bar{r}^{\prime },0\right)
=\prod_{j=1}^{N}K_{1}\left( {\bf r}_{j}^{\prime },\beta |{\bf r}%
_{j},0\right) .
\end{equation}%
The vector $\bar{r}$ denotes the configuration $\left( {\bf r}_{1},\ldots ,%
{\bf r}_{N}\right) ^{T}$ of the $N$ particles in the system. When fermions
in the same spin state are involved one has to anti-symmetrize this
propagator 
\begin{equation}
{\cal K}_{F}\left( \bar{r},\beta |\bar{r}^{\prime },0\right) =\frac{1}{N!}%
\sum_{P}\left( -1\right) ^{P}\prod_{j=1}^{N}K_{1}\left( \left\{ P{\bf r}%
\right\} _{j},\beta |{\bf r}_{j}^{\prime },0\right) {\bf .}
\end{equation}%
The summation runs over all permutations $P$ of the particle indices, and
the sign factor $\left( -1\right) ^{P}$ assures that odd permutations
contribute with a negative sign. For convenience in the notations, we limit
the treatment to the spin-polarized case. The non-polarized case can be
treated along the same lines, see e.g. \cite{LBD.PRE00}.

The partition function for this system of $N$ fermions is 
\[
{\cal Z}_{F}\left( \beta |N\right) =\int \int {\cal K}_{F}\left( \bar{r}%
,\beta |\bar{r}^{\prime },0\right) \delta \left( \bar{r}-\bar{r}^{\prime
}\right) d\bar{r}d\bar{r}^{\prime }. 
\]%
The integrals over the configuration space can be done using the
decomposition of the permutations in cycles  \cite%
{K.NH74,F.B72,F.PR53,M.PTP51,PO.PR56,G.JMP65,G.GB71,ML.PRA91,CM.PRC94}. With 
$M_{\ell }$ denoting the number of cycles of length $\ell ,$ the partition
function becomes 
\begin{eqnarray}
{\cal Z}_{F}\left( \beta |N\right) & = & \sum_{M_{1}\cdots M_{N}}\Theta \left(
\sum_{\ell =1}^{N}\ell M_{\ell }=N\right)\times \\ \nonumber
 &  & \prod_{\ell =1}^{N}\frac{\left(
-1\right) ^{\left( \ell -1\right) M_{\ell }}}{M_{\ell }!\ell ^{M_{\ell }}}%
\left( Z_{1}\left( \ell \beta \right) \right) ^{M_{\ell }}.
\end{eqnarray}
Introducing a generating function 
\begin{equation}
\Xi _{F}\left( \beta ,u\right) =\sum_{N=0}^{\infty }u^{N}{\cal Z}_{F}\left(
\beta |N\right) ,  \label{eq:genfunZF}
\end{equation}%
the constraint $\sum_{\ell =1}^{N}\ell M_{\ell }=N$ can be removed, and the
summation over $M_{\ell }$ can be done 
\begin{equation}
\Xi _{F}\left( \beta ,u\right) =\exp \left( -\sum_{\ell =1}^{\infty }\frac{%
\left( -u\right) ^{\ell }}{\ell }Z_{1}\left( \ell \beta \right) \right) .
\end{equation}%
Using the spectral representation (\ref{eq:Z1spectral}) of $Z_{1}\left(
\beta \right) ,$ the sum over the cycles can easily be performed. The result
is:\begin{eqnarray}
\Xi _{F}\left( \beta ,u\right)  & = & \exp \left( \int \Omega \left( E\right) \ln
\left( 1+ue^{-\beta E}\right) dE\right) \\ \nonumber
& = & \prod_{\nu =0}^{\infty }\left(
1+ue^{-\beta \epsilon _{\nu }}\right) ^{g_{\nu }}.
\end{eqnarray}
Again as in the case of a single particle, the two variables $\beta ,u$ have
no specific physical meaning with respect to the parameters defining the
system, i.e., the energy $E$ and the number of particles $N$. The partition
function ${\cal Z}_{F}\left( \beta |N\right) $ for $N$ particles can then be
obtained from its generating function $\Xi _{F}\left( \beta ,u\right) $ by
the inverse transformation%
\begin{equation}
{\cal Z}_{F}\left( \beta |N\right) =\frac{1}{N!}\left. \frac{d^{N}\Xi
_{F}\left( \beta ,u\right) }{du^{N}}\right| _{u=0}.
\end{equation}%
Remark that in the derivation of (\ref{eq:genfunZF}) $u$ is nothing but an
intermediate variable. If one substitutes $u$ by $1/z,$ one can state that $%
\Xi \left( \beta ,\frac{1}{z}\right) $ is the $Z$ transform of the sequence $$%
\left\{ 1,{\cal Z}_{F}\left( \beta |1\right) ,\ldots ,{\cal Z}_{F}\left(
\beta |N\right) ,\ldots \right\}. $$
The inverse of the $Z$ transform is
given by a contour-integration, leading to the following alternative
inversion formula: 
\begin{equation}
{\cal Z}_{F}\left( \beta |N\right) =\frac{1}{2\pi i}\oint_{C}\frac{\Xi
_{F}\left( \beta ,\gamma \right) }{\gamma ^{N+1}}d\gamma
\end{equation}%
where $C$ is a closed counterclockwise contour around the origin and chosen
in such a way that only the origin is a singularity. The inversion can also
be done using recursion formulas  \cite{S.PRC87,BF.JCP93}. Although they are
more appropriate for bosons, they are occasionally used for fermions  \cite%
{L.IP61}.

\subsection{The inversion method and the chemical potential}

Since $\Xi _{F}\left( \beta ,\gamma \right) $ is an analytic function of $%
\gamma ,$ a circular contour with radius $u$ can be considered, and
introducing $\gamma =ue^{i\theta }$ the inversion integral can be rewritten
as 
\begin{equation}
{\cal Z}_{F}\left( \beta |N\right) =\frac{1}{2\pi }\int_{0}^{2\pi }e^{\left[
\ln \Xi _{F}\left( \beta ,ue^{i\theta }\right) -N\ln u\right] }e^{-iN\theta
}d\theta .
\end{equation}%
Choosing the radius $u$ such that the argument 
$$\left[ \ln \Xi _{F}\left(
\beta ,ue^{i\theta }\right) -N\ln u\right] $$ of the exponential function
reaches an extremum, one obtains the condition 
\begin{equation}
N=u\frac{d}{du}\ln \Xi _{F}\left( \beta ,u\right) =\int \Omega \left(
E\right) \frac{ue^{-\beta E}}{1+ue^{-\beta E}}dE.  \label{eq:uext}
\end{equation}%
Denoting the solution $u$ obtained from this implicit equation by $u_{e}$,
the partition function can be rewritten as 
\begin{eqnarray}
{\cal Z}_{F}\left( \beta |N\right) &=&{\cal Z}_{e}\left( \beta |N\right)
\int_{0}^{2\pi }\Psi \left( \theta \right) d\theta , \\
{\cal Z}_{e}\left( \beta |N\right) &=&\frac{\Xi \left( \beta
,u_{e}\right) }{u_{e}}, \\
\Psi \left( \theta \right) &=&\frac{1}{2\pi }\frac{\Xi \left(
\beta ,u_{e}e^{i\theta }\right) }{\Xi \left( \beta ,u_{e}\right) }%
e^{-iN\theta }.
\end{eqnarray}%
The free energy ${\cal F}_{F}\left( \beta |N\right) =-\frac{1}{\beta }\ln 
{\cal Z}_{F}\left( \beta |N\right) $ therefore consists of two contributions 
\begin{equation}
{\cal F}_{F}\left( \beta |N\right) ={\cal F}_{e}\left( \beta |N\right) -%
\frac{1}{\beta }\ln \left( \int_{0}^{2\pi }\Psi \left( \theta \right)
d\theta \right) .  \label{eq:Fexact}
\end{equation}%
\begin{equation}
\rm\rm{with }{\cal F}_{e}\left( \beta |N\right) =-\frac{1}{\beta }\ln \frac{%
\Xi \left( u_{e},\beta \right) }{u_{e}^{N}},  \label{eq:Fappr}
\end{equation}%
If one would redefine $u_{e}$ as $u_{e}=e^{\beta \mu },$ the relation (\ref%
{eq:uext}) becomes the common expression for the chemical potential $\mu $
given the total number of particles $N$. However, this expression is a
consequence of the mathematical technique used for the inversion of the
generating function, and it leads to an {\em approximation} for the
partition function.

It is well known that the steepest descent method often gives good results
in the case of inverse Laplace transforms, but several suggestions are found
in the literature  \cite{HK.AnP99,L.IP61,S.CUP46,D.PCPS49,D.AP73} to use this
method also for this type of inversion of a $Z$ transform. In this case, the
approximation turns out to be rather poor, and in general the error estimate
poses a problem. Below we shall show that, although the approximation (\ref%
{eq:Fappr}) is excellent for fermion systems with completely filled shells
in the low temperature limit, it requires important corrections when the
shells are not completely filled. It should be emphasized that there is no
approximation involved in the use of $u_{e}$ (or any other integration
radius $u)$ as long as only the singularity in the origin is contained in
the integration domain. The approximation stems from the replacement of the
integrand by its extremum, in the hope that the contribution of the
remaining oscillatory contribution to the free energy is negligible.

\subsection{The modified fugacity}

The optimal radius $u_{e}$ for inversion can be calculated from (\ref%
{eq:uext}) if the energy spectrum of the system is known. After integration over 
$E,$ one obtains the following transcendental equation for $u_{e}:$ 
\begin{equation}
N=\sum_{\nu }g_{\nu }\frac{u_{e}e^{-\beta \epsilon _{\nu }}}{%
1+u_{e}e^{-\beta \epsilon _{\nu }}}.
\end{equation}%
Assuming the usual relation between the chemical potential and the fugacity $%
u_{e}=e^{\beta \mu }$ allows in the low temperature limit only solutions
that consist of closed shells \cite{SW.PRA98}. In order to have the
possibility to incorporate also open shells we assume that $u_{e}$ depends
on $\beta $ in the low temperature limit as follows: 
\begin{equation}
u_{e}=\lambda e^{\beta \epsilon _{L}},
\end{equation}%
where $L$ is the {\em lowest not fully occupied} level. Splitting the
summation over the levels in the equation for $N$ into three parts we obtain 
\begin{eqnarray}
N & = & \sum_{\nu =0}^{L-1}g_{\nu }\frac{\lambda e^{\beta \left( \epsilon
_{L}-\epsilon _{\nu }\right) }}{1+\lambda e^{\beta \left( \epsilon
_{L}-\epsilon _{\nu }\right) }} \\ \nonumber
 & + & g_{L}\frac{\lambda }{1+\lambda }+\sum_{\nu
=L+1}^{\infty }g_{\nu }\frac{\lambda e^{\beta \left( \epsilon _{L}-\epsilon
_{\nu }\right) }}{1+\lambda e^{\beta \left( \epsilon _{L}-\epsilon _{\nu
}\right) }}, 
\end{eqnarray}
and taking the limit $\beta \rightarrow \infty $ we find the following
implicit expression for the parameter $\lambda $: 
\begin{equation}
N=N_{L}+g_{L}\frac{\lambda }{1+\lambda },
\end{equation}%
where $N_{L}$ is the total number of states in all fully occupied levels 
\begin{equation}
N_{L}=\sum_{\nu =0}^{L-1}g_{\nu }.  \label{eq:N_L}
\end{equation}%
This number and therefore the level $L$ itself, has to be determined from 
\begin{equation}
N_{L}\leq N<N_{L+1}.
\end{equation}%
Once $N_{L}$ is known, one can calculate the value of $\lambda ,$ i.e. 
\begin{equation}
\lambda =\frac{N-N_{L}}{N_{L}+g_{L}-N}=\frac{N-N_{L}}{N_{L+1}-N}.
\end{equation}%
Note that this calculation explicitly assumed that $N$ differs from $N_{L}$.
In the case that $N=N_{L}$, i.e., if the level $L$ is empty, one obtains $%
u_{e}=e^{\beta \epsilon _{L}}$ in the low temperature limit. It is
interesting to remark that the divergent terms in the steepest descent
inversion for bosons could be eliminated by Landsberg using a similar
modification of the fugacity  \cite{L:PR54}.

Using now the generating function $\Xi \left( \beta ,u_{e}\right) $ in the
limit $\beta \rightarrow \infty $ with $u_{e}=\lambda e^{\beta \epsilon
_{L}},$ one obtains the following zero-order approximation ${\cal Z}%
_{e}\left( \beta |N\right) $ for the partition function 
\begin{equation}
\lim_{\beta \rightarrow \infty }{\cal Z}_{e}\left( \beta |N\right) =\frac{e^{-\beta E_{0}}\left( N_{L+1}-N_{L}\right) ^{N_{L+1}-N_{L}}}{\left( N-N_{L}\right)
^{N-N_{L}}\left( N_{L+1}-N\right) ^{N_{L+1}-N}}
\end{equation}
with the fermion ground state energy $E_{0}$ given by 
\begin{equation}
E_{0}=\sum_{\nu =0}^{L-1}g_{\nu }\epsilon _{\nu }+\left( N-N_{L}\right)
\epsilon _{L}.  \label{eq:E_0}
\end{equation}
The integral correction $\int_{0}^{2\pi }\Psi \left( \theta \right) d\theta $
to the partition function can be derived from 
\begin{equation}
\lim_{\beta \rightarrow \infty }\Psi \left( \theta \right) =\frac{1}{2\pi }%
\left( \frac{1+\lambda e^{i\theta }}{1+\lambda }\right) ^{g_{L}}e^{-i\theta
\left( N-N_{L}\right) },
\end{equation}
and hence, using the binomial expansion of $\left( 1+\lambda e^{i\theta
}\right) ^{g_{L}}$, one obtains 
\begin{eqnarray}
\lim_{\beta \rightarrow \infty }\int_{0}^{2\pi }\Psi \left( \theta \right)
d\theta & = & {N_{L+1}-N_{L} \choose N-N_{L}}\left( N-N_{L}\right) ^{N-N_{L}} \\ \nonumber
 & \times & \frac{\left( N_{L+1}-N\right) ^{N_{L+1}-N}}{%
\left( N_{L+1}-N_{L}\right) ^{N_{L+1}-N_{L}}} 
\end{eqnarray}

The partition function thus becomes 
\begin{equation}
\lim_{\beta \rightarrow \infty }{\cal Z}_{F}\left( \beta |N\right) {\
\rightarrow }%
{g_{L} \choose N-N_{L}}%
e^{-\beta E_{0}},  \label{eq:Zresult}
\end{equation}
where the combinatorial factor in front accounts for the number of
possibilities to occupy the open shell. In order to obtain this factor, the
modification of the fugacity is not sufficient: the inversion procedure has
to be carried out completely. In other words: the open shell partition
function of fermions in the low temperature limit has to be calculated by an
inversion because the contribution of the integral correction turns out to
be crucial!

But the exact ground state energy (\ref{eq:E_0}) is obtained from the
approximate expression (\ref{eq:Fappr}). This means that the integral
correction to the free energy (\ref{eq:Fexact}) becomes negligible in the
low temperature limit; it only contributes to the degeneracy factor in the
partition function, i.e., to the entropy. In the next section we will
investigate whether this coincidence also remains valid for the correlation
functions.

\section{The density of an open shell Fermion gas}

In this section we derive the one-point correlation function that describes
the density in the ground-state, using the exact inversion. Furthermore, we
show that the use of the modified fugacity recovers the exact result.

\subsection{The density and its generating function}

The generating function for the Fourier transform $\tilde{n}_{{\bf q}}$ of the density of a $N$-fermion system  has been derived in  \cite{BDL.PRE98a}%
. It is given by 
\begin{equation}
{\cal G}_{\tilde{n}_{{\bf q}}}\left( \beta ,u\right) =\sum_{N=0}^{\infty }%
\tilde{n}_{{\bf q}}\left( \beta ,N\right) {\cal Z}_{F}\left( \beta |N\right)
Nu^{N},  \label{eq:generate_n}
\end{equation}%
and it can easily be expressed in terms of the density of states 
\begin{equation}
\frac{{\cal G}_{\tilde{n}_{{\bf q}}}\left( \beta ,u\right) }{\Xi _{F}\left(
\beta ,u\right) }{\bf =}\int f_{E}\left( u,\beta \right) \Omega \left( E,%
{\bf r}\right) e^{-i{\bf q\cdot r}}d{\bf r}dE{\bf ,}
\label{eq:generate_n_densstates}
\end{equation}%
where $f_{E}\left( u,\beta \right) $ has the form of an occupation function,
but $u$ is still the auxiliary variable to construct the power series 
\begin{equation}
f_{E}\left( u,\beta \right) =\frac{ue^{-\beta E}}{1+ue^{-\beta E}}.
\label{eq:occupationfunction}
\end{equation}

\subsection{The exact inversion}

Inverting (\ref{eq:generate_n}) for $\tilde{n}_{{\bf q}}\left( \beta
,N\right) $ by the contour integral technique, and filling out ${\cal G}_{%
\tilde{n}_{{\bf q}}}\left( \beta ,z\right) $ in the form (\ref%
{eq:generate_n_densstates})%
\begin{eqnarray}
\tilde{n}_{{\bf q}}\left( \beta ,N\right)  & = & \frac{1}{N}\frac{1}{{\cal Z}%
_{F}\left( \beta |N\right) }\sum_{\nu =0}^{\infty }n_{\nu }\left( {\bf q}%
\right) \frac{1}{2\pi i} \\ \nonumber
 & \times & \lim_{u\rightarrow 0}\oint\limits_{\left| z\right|
=u}\Xi _{F}\left( \beta ,z\right) \frac{ze^{-\beta \epsilon _{\nu }}}{%
1+ze^{-\beta \epsilon _{\nu }}}\frac{dz}{z^{N+1}},
\end{eqnarray}
the contour integral can be transformed into an angular integral. One
obtains after some effort 
\begin{equation}
\lim_{\beta \rightarrow \infty }\tilde{n}_{{\bf q}}\left( \beta ,N\right)
\rightarrow \frac{1}{N}\left( \sum_{\nu =0}^{L-1}n_{\nu }\left( {\bf q}%
\right) +\frac{N-N_{L}}{N_{L+1}-N_{L}}n_{L}\left( {\bf q}\right) \right) ,
\label{eq:density}
\end{equation}%
with $n_{\nu }\left( {\bf q}\right) $ is defined as the Fourier transform of
the density contribution from the single-particle wave functions in the
level $\nu $%
\begin{equation}
n_{\nu }\left( {\bf q}\right) =\sum_{k=1}^{g_{\nu }}\int \left| \varphi
_{\nu ,k}\left( {\bf r}\right) \right| ^{2}e^{-i{\bf q\cdot r}}d{\bf r.}
\end{equation}%
This result for the density could have been guessed by realizing that for an
open shell the states belonging to that shell must have the same probability
of being occupied.

\subsection{The optimal fugacity}

The generating function for the density in the spectral
representation (\ref{eq:Omegadiag}) can be written as 
\begin{equation}
\frac{{\cal G}_{\tilde{n}_{{\bf q}}}\left( \beta ,u\right) }{\Xi _{F}\left(
\beta ,u\right) }{\bf =}\sum_{\nu =0}^{\infty }\frac{ue^{-\beta \epsilon
_{\nu }}}{1+ue^{-\beta \epsilon _{\nu }}}n_{\nu }\left( {\bf q}\right) .
\end{equation}%
Repeating the {\em approximate inversion} as for the partition function from
the previous section, using again 
$$u_{e}=\lambda e^{\beta \epsilon _{L}},$$
one finds 
\begin{equation}
\lim_{\beta \rightarrow \infty }\frac{{\cal G}_{\tilde{n}_{{\bf q}}}\left(
\beta ,u_{e}\right) }{\Xi _{F}\left( \beta ,u_{e}\right) }=\sum_{\nu
=0}^{L-1}n_{\nu }\left( {\bf q}\right) +\frac{N-N_{L}}{N_{L+1}-N_{L}}%
n_{L}\left( {\bf q}\right) .
\end{equation}%
The modified fugacity thus allows to calculate the ground-state density
exactly, similarly as we found for the ground-state energy.

\subsection{A useful relation}

Because the generating function often can be written in closed form, the
following exact relation 
\begin{equation}
\lim_{\beta \rightarrow \infty }\tilde{n}_{{\bf q}}\left( \beta ,N\right) =%
\frac{1}{N}\lim\limits_{\beta \rightarrow \infty }\frac{{\cal G}_{\tilde{n}_{%
{\bf q}}}\left( \beta ,u_{e}\right) }{\Xi _{F}\left( \beta ,u_{e}\right) },
\label{eq:numdens}
\end{equation}%
between the results for the ground-state of the two inversion methods turns
out to be of practical numerical importance. The density, i.e. the sum of
single-particle contributions over all the occupied states, can be obtained
from an appropriate limit of a ratio of two generating functions. For
fermions in a harmonic potential ${\cal G}_{\tilde{n}_{{\bf q}}}\left( \beta
,u\right) /\Xi _{F}\left( \beta ,u\right) $ is known in closed form and the
limit can be used to obtain an explicit result for the density, thus
avoiding a tedious summation over occupied states.

\section{The polarization function of an open shell fermion gas}

In this section we investigate whether the two-point correlation function that describes
the density-density correlations of the fermion system can also be obtained in the low temperature limit using a modified chemical potential.
Unfortunately, this is not the case. The exact inversion of the generating function and the use of the chemical potential give analogous but slightly different results. Both calculations  use the same techniques as explained in the preceding sections. Therefore only an outline of the method and a summary of the results will be given.

The Fourier transform of the two-point correlation function is defined as:

\begin{equation}
N^{2}\tilde{C}_{{\bf q^{\prime },q}}\left( \tau |\beta ,N\right) =%
\left\langle \sum_{j=1}^{N}\sum_{l=1}^{N}e^{-i{\bf q}^{\prime }{\bf \cdot r}_{j}^{\prime }\left( \tau \right) }e^{-i{\bf q\cdot r}_{l}\left( 0\right)
}\right\rangle.
\end{equation}%
An outline of the derivation of this function can be found in  \cite%
{LSBD.SSC01}. Defining the following generating function for the two point
correlation function 
\begin{equation}
G_{\tilde{C}_{{\bf q^{\prime },q}}}\left( \tau |\beta ,u\right)
=\sum_{N=0}^{\infty } \tilde{C}_{{\bf q^{\prime },q}}\left( \tau
|\beta ,N\right) {\cal Z}_{F}\left( \beta |N\right) N^{2}u^{N},
\end{equation}%
a straightforward application of the techniques illustrated in our former
publications  \cite{BDL.PRE97a,BDL.PRE98a} leads after a lengthy but straightforward calculation to an expression obtained by  performing the contour
integral of the inversion exactly in the limit $\beta \rightarrow \infty $ : 
\begin{eqnarray} \label{eq:twopcexact}
\lim_{\beta \rightarrow \infty }N^{2}\tilde{C}_{{\bf q^{\prime },q}}\left(
\tau |\beta ,N\right) &=&A_{0}\left( {\bf q,q}^{\prime },\tau \right)\\ \nonumber  &+&\frac{%
N-N_{L}}{N_{L+1}-N_{L}}A_{1}\left( {\bf q,q}^{\prime },\tau \right) 
 \\ \nonumber
&+&\frac{N-N_{L}}{N_{L+1}-N_{L}}\frac{N-N_{L}-1}{N_{L+1}-N_{L}-1}\\ \nonumber &\times& A_{2}\left( 
{\bf q,q}^{\prime },\tau \right) , 
\end{eqnarray}%
with 
\begin{eqnarray}
A_{0}\left( {\bf q,q}^{\prime },\tau \right) &=&\sum\limits_{\nu
=0}^{L-1} n_{\nu }\left( {\bf q}\right) \sum\limits_{\nu ^{\prime
}=0}^{L-1}n_{\nu ^{\prime }}\left( {\bf q}^{\prime }\right) \\ \nonumber
&+&\sum\limits_{\nu
=0}^{L-1}\sum\limits_{\nu ^{\prime }=L}^{\infty }e^{\tau \left( \epsilon _{\nu
}-\epsilon _{\nu ^{\prime }}\right) }M_{\nu ,\nu ^{\prime }}\left( {\bf q,q}%
^{\prime }\right)  , \\
A_{1}\left( {\bf q,q}^{\prime },\tau \right) &=&n_{L}\left( {\bf q}\right)
\sum\limits_{\nu =0}^{L-1}n_{\nu }\left( {\bf q}^{\prime }\right)\\ \nonumber
&+& n_{L}\left( {\bf q}^{\prime }\right) \sum\limits_{\nu =0}^{L-1}n_{\nu
}\left( {\bf q}\right)  \\ \nonumber
&&+\sum\limits_{\nu =L}^{\infty }e^{\tau \left( \epsilon _{L}-\epsilon _{\nu
}\right) }M_{L,\nu }\left( {\bf q,q}^{\prime }\right) \\ \nonumber &-&\sum\limits_{\nu
=0}^{L-1}e^{\tau \left( \epsilon _{\nu }-\epsilon _{L}\right) }M_{\nu
,L}\left( {\bf q,q}^{\prime }\right) , \\
A_{2}\left( {\bf q,q}^{\prime },\tau \right) &=&n_{L}\left( {\bf q}\right)
n_{L}\left( {\bf q}^{\prime }\right) -M_{L,L}\left( {\bf q,q}^{\prime
}\right) ,
\end{eqnarray}%
and
\begin{eqnarray}
M_{\nu ,\nu ^{\prime }}\left( {\bf q,q}^{\prime }\right)
&=&\sum_{k=1}^{g_{\nu }}\sum_{k^{\prime }=1}^{g_{\nu ^{\prime }}}\Lambda
_{\nu ,k;\nu ^{\prime },k^{\prime }}\left( {\bf q}\right) \Lambda _{\nu
^{\prime },k^{\prime };\nu ,k}\left( {\bf q}^{\prime }\right) , \\
\Lambda _{\nu ,k;\nu ^{\prime },k^{\prime }}\left( {\bf q}\right) &=&\int
\varphi _{\nu ,k}^{\ast }\left( {\bf r}\right) \varphi _{\nu ^{\prime
},k^{\prime }}\left( {\bf r}\right) e^{-i{\bf q\cdot r}}d{\bf r}.
\end{eqnarray}

It should be noted that the expression for $A_{2}\left( {\bf q,q}^{\prime
},\tau \right) $ implies that this term is independent of Euclidean time $%
\tau $.

It turns out that the result (\ref{eq:twopcexact}) is {\em not }obtained
with the approximate inversion $G_{\tilde{C}_{{\bf q^{\prime },q}}}\left(
\tau |\beta ,u_{e}\right) /\Xi _{F}\left( \beta ,u_{e}\right) $ of the
generating function. Using again $u_{e}=\lambda e^{\beta \epsilon _{L}}$ the
dominant terms for $\beta \rightarrow \infty $ instead become 
\begin{eqnarray}
\lim_{\beta \rightarrow \infty }\frac{G_{\tilde{C}_{{\bf q^{\prime },q}%
}}\left( \tau |\beta ,u_{sd}\right) }{\Xi _{F}\left( \beta ,u_{sd}\right) }
&=&A_{0}\left( {\bf q,q}^{\prime },\tau \right) \\ \nonumber &+&\frac{N-N_{L}}{N_{L+1}-N_{L}%
}A_{1}\left( {\bf q,q}^{\prime },\tau \right)  \\ \nonumber
&+&\frac{N-N_{L}}{N_{L+1}-N_{L}}\frac{N-N_{L}}{N_{L+1}-N_{L}}\\ \nonumber &\times& A_{2}\left( 
{\bf q,q}^{\prime },\tau \right) .  \label{eq:twopointapp}
\end{eqnarray}%
Although very similar to the exact result (\ref{eq:twopcexact}) the
combinatorial factors for the not fully occupied level are slightly
different. The justification of this approximate inversion is rather
elaborate. E.g., it is not trivial that the contour integration around the
origin can be extended to the large radius $u_{e}$ without introducing extra
poles. This point  has been carefully checked.

Although very similar to the exact result (\ref{eq:twopcexact}) the
combinatorial factors in (\ref{eq:twopointapp}) for the not fully occupied
level are not correct. Therefore the similar correspondence as found for the
partition function and for the density (\ref{eq:numdens}), is valid for the two-point
correlation function in the case of{\em \ closed shells only}. Indeed, $%
A_{1}\left( {\bf q,q}^{\prime },\tau \right) $ and $A_{2}\left( {\bf q,q}%
^{\prime },\tau \right) $ do not contribute to the polarization function for
 the case of closed shells, i.e., for $N=N_{L}$. This leads  to:
\begin{equation}
\lim_{\beta \rightarrow \infty }N^{2}\tilde{C}_{{\bf q^{\prime },q}}\left(
\tau |\beta ,N\right) =\lim_{\beta \rightarrow \infty }\frac{G_{\tilde{C}_{%
{\bf q^{\prime },q}}}\left( \tau |\beta ,u_{e}\right) }{\Xi \left( \beta
,u_{e}\right) }.
\end{equation}

\section{Discussion and Conclusions}

Several authors suggested or stated that calculations performed with the
methods developed in the grand canonical ensemble cannot be trusted when one
deals with a given finite number of particles  \cite%
{ZUK.PRp77,FHW.JSP70,BDL.SSC96,GH.PRL97,NBGIR.PRL97,GR.PRL97,IGNWR.PRL99,HK.AnP99,R.PRC95,P.PRL00,MS:PRL00}%
. In this paper we have pinpointed the origin of this discrepancy for open
shell Fermion systems. Our analysis is based on an exact inversion of the
existing transform between the partition function, the one-point correlation
function and the two point-correlation function for $N$ particles and their
generating functions. We observed that the generating function of the
partition function for the proper parameter choice becomes the grand
canonical partition function of the same model. This observation allows to
compare the results for the exact inversion with those relying on the
chemical potential. It should be stressed that a modification of the
relation between the fugacity and the chemical potential is needed to obtain
the correct ground state energy, but this modification is {\em not sufficient%
} to obtain the low temperature limit of the partition function correctly in
the case of open shell fermion gases: the entropy requires the exact
inversion. In summary the modification of the fugacity definition for open
shells leads to the correct expressions for the ground-state energy and the
ground-state density. But it is not good enough to obtain the low temperature
limit of the partition function and the polarization function, and a
correction is needed, as derived above. This situation is summarized in
table 1.
\begin{table}[tbp] \centering%
\caption{This table summarizes for which quantities and type of systems the use
 of the unmodified chemical potential $\mu$ and of its modified form leads to correct 
results.\label{table}}
\begin{tabular}{lll}
\hline\noalign{\smallskip}
{\bf Quantity} & ${\bf \mu }$ & {\bf Modified }$\mu $ \\
\noalign{\smallskip}\hline\noalign{\smallskip}
Ground state energy & closed shell & open shell \\ 
Density & closed shell & open shell \\ 
Partition function & closed shell & extra correction \\ 
Correlation function & closed shell & extra correction \\
\hline\noalign{\smallskip}
\end{tabular}%
\end{table}

Note that the differences discussed here have their origin in the degeneracy
of the lowest not completely occupied level. In the absence of this
degeneracy, i.e. for closed shell systems, the ground-state expressions of
both inversion schemes are identical. But if there are $N-N_{L}$ fermions in
the lowest not completely occupied level, a correction term proportional to$$%
\frac{N-N_{L}}{N_{L+1}-N_{L}}$$is introduced in the ground-state energy and
the density by modifying the fugacity. An extra term proportional to $$\frac{%
(N-N_{L})(N-N_{L}-1)}{(N_{L+1}-N_{L})(N_{L+1}-N_{L}-1)}$$ is found for the polarization
function.

The relation between the density of states, the partition function and the
generating function have been discussed by several authors. In classical
statistical mechanics they are attributed to Fowler and Darwin  \cite%
{DF:PM22,DF.PCPS23}. In quantum statistical mechanics they are discussed in
the book by P.T. Landsberg  \cite{L.IP61} who makes an epistemological
analysis of them (see, e.g., ref.  \cite{Sn01}). A mathematical justification
and review of the present understanding can be found in  \cite%
{KostrykinSchrader}. Of course the present findings also oblige to
scrutinize the use of the chemical potential in other calculation schemes
like the density functional approach or the Gross-Pitaevsky equation. The
questions are there: ``How can one control these methods, clearly based on
the grand canonical ensemble, on their correctness for a finite number of
particles?'', and ``Is a modification of the chemical potential in these
cases also necessary?''. Our preliminary investigations indicate that in the case that the levels are degenerate, one should use the modified chemical potential in order to get the energy and the density correct, and to obtain a fair approximation for the two-point correlation function.

\begin{acknowledgement}
This work is performed within the framework of the FWO
projects No. 1.5.545.98, G.0287.95, G.0071.98,\\ and WO.073.94N
[Wetenschappelijke Onderzoeksgemeenschap van het FWO over ``Laagdimensionele
systemen'' (Scientific Research Community on Low-Dimensional Systems)], the
``Interuniversitaire Attractiepolen -- Belgische Staat, Diensten van de
Eerste Minister -- Wetenschappelijke, Technische en Culturele
Aangelegenheden'' (Interuniversity Poles of Attraction Programs --Belgian
State, Prime Minister's Office --Federal Office for Scientific, Technical
and Cultural Affairs), and in the framework of the GOA BOF\ UA 2000 projects
of the Universiteit Antwerpen.
\end{acknowledgement}
%
%

\end{document}